\documentclass[12pt,preprint]{aastex}
\usepackage{txfonts}
\usepackage{graphicx}
\usepackage{amssymb}
\usepackage{epsfig}
\usepackage{natbib}
\usepackage{textcomp}

\begin{document}

\title{Small-scale structure of the interstellar medium towards $\rho$~Oph stars: diffuse band observations}

\author{M. A. Cordiner\altaffilmark{1,}\altaffilmark{2}, S. J. Fossey\altaffilmark{3}, A. M. Smith\altaffilmark{1,}\altaffilmark{4} and P. J. Sarre\altaffilmark{1}}
\email{martin.cordiner@nasa.gov}

\altaffiltext{1}{School of Chemistry, The University of Nottingham, University Park, Nottingham, NG7 2RD, United Kingdom}
\altaffiltext{2}{Current Address: Astrochemistry Laboratory and The Goddard Center for Astrobiology, Mailstop 691, NASA Goddard Space Flight Center, 8800 Greenbelt Road, Greenbelt, MD 20770, USA}
\altaffiltext{3}{Department of Physics and Astronomy, University College London, Gower Street, London, WC1E 6BT, United Kingdom}
\altaffiltext{4}{Current Address: Adler Planetarium, 1300 South Lake Shore Drive, Chicago, IL 60605, USA}

\shorttitle{\sc Small-scale structure of the ISM towards $\rho$~Oph stars: DIB observations}
\shortauthors{\sc Cordiner, Fossey, Smith and Sarre}

\begin{abstract}
We present an investigation of small-scale-structure in the distribution of large molecules/dust in the interstellar medium through observations of diffuse interstellar bands (DIBs). High signal-to-noise optical spectra were recorded towards the stars $\rho$~Oph A, B, C and DE using the University College London Echelle Spectrograph (UCLES) on the Anglo-Australian Telescope. The strengths of some of the DIBs are found to differ by about $5-9$\% between the close binary stars $\rho$~Oph A and B, which are separated by a projected distance on the sky of only \emph{c.} 344 AU. This is the first star-system in which such small-scale DIB strength variations have been reported. The observed variations are attributed to differences between a combination of carrier abundance and the physical conditions present along each sightline. The sightline towards $\rho$~Oph C contains relatively dense, molecule-rich material and has the strongest $\lambda\lambda5850$ and 4726 DIBs. The gas towards DE is more diffuse and is found to exhibit weak `C$_2$' (blue) DIBs and strong yellow/red DIBs. The differences in diffuse band strengths between lines of sight are, in some cases, significantly greater in magnitude than the corresponding variations among atomic and diatomic species, indicating that the DIBs can be sensitive tracers of interstellar cloud conditions.
 
\end{abstract}

\keywords{ISM: lines and bands --- ISM: clouds --- ISM: structure --- stars: individual ($\rho$~Oph A, B, C and DE)}

\section{Introduction}

Knowledge of the spatial distribution, dynamics, composition and chemistry of neutral and ionized gas and dust is of fundamental importance in understanding the interstellar medium (ISM) and its relation to star formation. Through observations of absorption lines of atoms and small molecules at wavelengths ranging from radio to ultraviolet, the gaseous component of diffuse and translucent interstellar clouds has been probed in some detail and indicates that filamentary or sheet-like small-scale-structure is common on scales as small as a few AU \citep{wat96,har03,hei07}. In contrast, the distribution of large molecules and dust in the diffuse ISM is far less well characterized, although for denser regions insight has been gained through imaging of dust emission. Variations in diffuse interstellar band strengths have previously been reported on distance scales down to a few parsecs \citep{poi04,van09}, but over distance scales $\sim10-150$~AU, no significant variations have yet been observed \citep{smi12,rol03}. The results presented in the present study constitute the first detection of DIB variations over distances on the order of a few hundred AU.
   
Although the carriers of the DIBs remain to be identified \citep{her95,sar06}, it is widely thought that the absorption features arise from electronic transitions in large gas-phase molecules which are associated with, and are possibly formed from, dust grains.  The linking of diffuse band absorption to the presence of dust grains rests in significant part on the generally good correlation between their strengths and $E_{B-V}$ although deviations are well established.

\begin{figure}
\centering
\includegraphics[width=0.7\columnwidth]{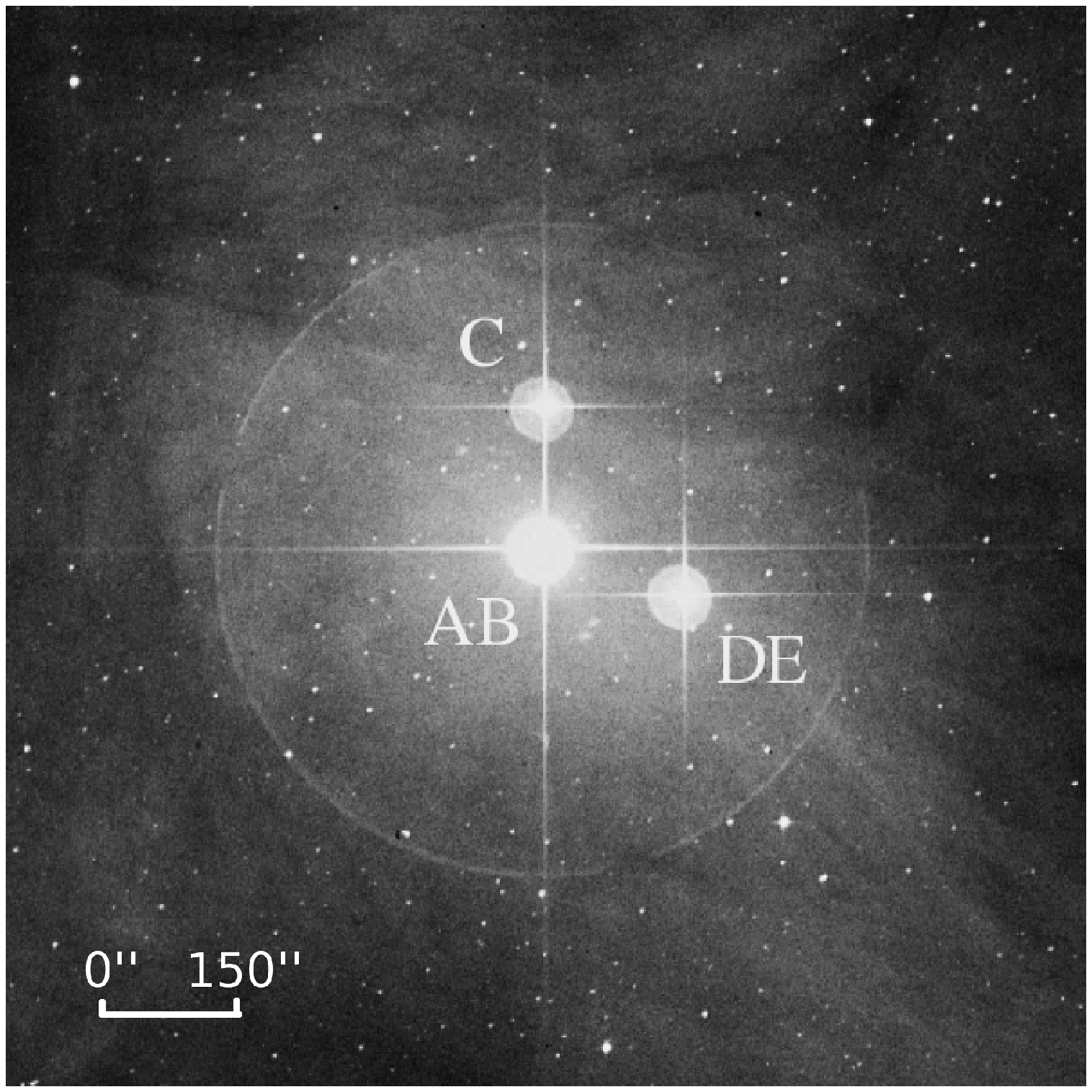}
\caption{Photographic $J$ (blue) image of the $\rho$ Oph system obtained from the Digitized Sky Survey (DSS). Angular distances between components are as follows: $\theta({\rm AB})=3.1''$, $\theta({\rm AC})=150''$, $\theta({\rm AD})=169''$, $\theta({\rm DE})=0.6''$. \label{fig:stars}}
\end{figure}

In order to probe the large molecule/dust distribution in diffuse regions of Ophiuchus we have undertaken observations of diffuse interstellar bands towards four bright stars in the  $\rho$~Oph system (A, B, C and DE) -- as shown in Figure \ref{fig:stars} -- the first results of which were described by \citet{cor05,cor06}. Taking a distance to $\rho$~Oph~A and B of 111$^{+12}_{-10}$~pc \citep{van07}, the stars comprising this binary are separated by a sky-projected distance of \emph{c.}~344~AU. The stars $\rho$~Oph~C and DE are more distant at 125$^{+14}_{-11}$ and 135$^{+12}_{-10}$~pc \citep{van07}, and separated from $\rho$~Oph~AB by sky-projected distances $\sim$~17,000~AU and $\sim$ 19,000~AU, respectively.  An additional motivation for this work is investigation of the relative band strengths for closely aligned lines-of-sight as this could provide a new clue in the search for diffuse band assignments. Of particular importance in this regard, the physical and chemical conditions of very closely-separated sightlines are expected to be similar to each other, thereby reducing the number of physical and chemical variables that may affect diffuse band strengths. In this \emph{Letter} we describe the detection of significant differences in DIB strengths of up to \emph{c.} 9\% between the closely-spaced lines of sight towards $\rho$~Oph~A and B, and larger differences up to \emph{c.} 40\% between AB and $\rho$~Oph~C and DE, all four of which have the same $E_{B-V}$ within $\pm$~0.01.

\section{Observations}

The $\rho$~Oph system consists of five young early-to-mid B-type dwarf stars \citep{dom94}, basic properties of which are given in Table \ref{tab:stars}. Observations were carried out during the months of 2004 March and 2005 April using UCLES at the Anglo-Australian Telescope with a $1''$ slit width. Complete wavelength coverage was obtained from $4,600-10,000$~\AA, with a spectroscopic resolving power of $52,000-58,000$. The median seeing was $0.95''$ (2004) and $1.7''$ (2005). This was sufficient for the $\rho$~Oph AB pair to be resolved on both epochs, with at most 0.1\% cross-contamination of flux in their reduced spectra (assuming Gaussian spatial profiles). The separation of $\rho$~Oph DE is only 0.6$''$, but D dominates the spectrum because it is 1.2~mag brighter than E.

The search for small-scale-structure through diffuse interstellar band spectra is facilitated where there are similarities in stellar spectral type coupled with a low level of contaminating stellar photospheric features.  The spectral match and stellar spectra are particularly good in the case of $\rho$~Oph A and B, and the situation is generally satisfactory for comparison of AB with $\rho$~Oph~DE. The spectrum of $\rho$~Oph~C is contaminated by weak absorption features that are not present for the other stars. Short-timescale variability is detected in several of its Si\,\textsc{ii} and He\,\textsc{i} lines, somewhat reminiscent of $\beta$~Cephei-type or slowly-pulsating B stars \citep[see, \emph{e.g.}][]{tel06}. Contamination of the observed $\rho$~Oph~C DIBs is negligible however, with the exception of $\lambda\lambda5780$ and 5797.

Reduction and analysis techniques have been described previously in detail by \citet{cor06}. Particular attention has been paid to non-linearity in the CCD response, scattered light subtraction, telluric and blaze corrections. All four program stars were observed in immediate succession and without altering the optical setup of the telescope or spectrograph (during each night of observations), in order to maximise consistency between the recorded spectra. Exposure times were varied so as to obtain similar photon counts for each star; signal-to-noise ratios of the reduced spectra are given in Table \ref{tab:stars}.

\section{Results and Discussion}

\subsection{Spectral observations}

\begin{figure}
\centering
\includegraphics[width=0.55\columnwidth]{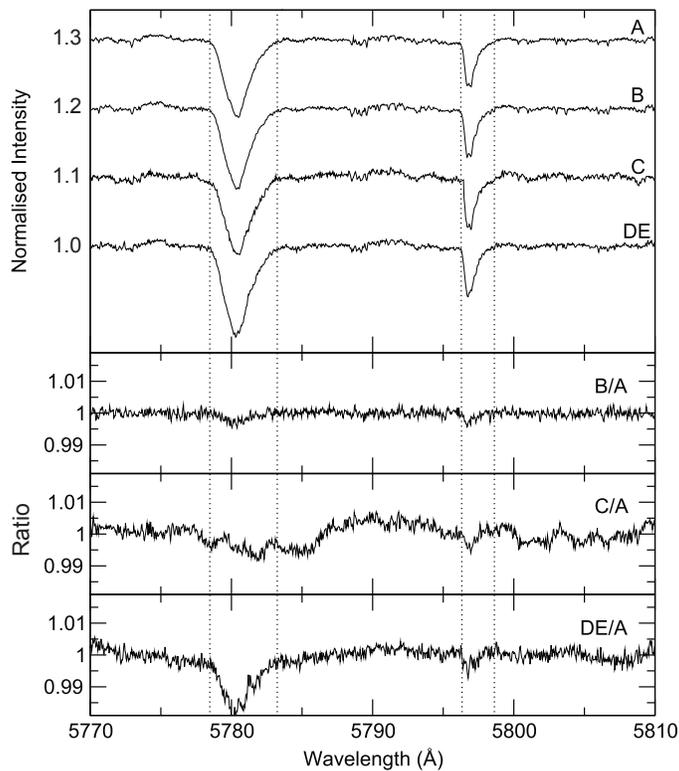}
\caption{Spectra of the region covering the $\lambda\lambda$5780 and 5797 diffuse bands.  Top four panels are observations towards $\rho$~Oph A, B, C and DE, with ordinate offsets of 0.1; these are corrected for the blaze function but not for telluric effects and are unsmoothed.  The three lower panels are  non-blaze-corrected non-telluric-corrected ratios I$_\lambda$(B)/I$_\lambda$(A), I$_\lambda$(C)/I$_\lambda$(A) and I$_\lambda$(DE)/I$_\lambda$(A).  The spectrum of $\rho$~Oph C is significantly contaminated by weak stellar lines in this region. For a ratio X/A, a value $<$~1 means the band is stronger along sightline X than A. \label{fig:5780}}
\end{figure}

Figure \ref{fig:5780} shows the reduced (non-telluric-corrected) spectra (I$_\lambda$ \emph{vs.} $\lambda$) in the wavelength region covering the $\lambda\lambda$5780 and 5797 diffuse interstellar bands towards $\rho$~Oph A, B, C and DE. The lower three panels show the ratios: I$_\lambda$(B)~/~I$_\lambda$(A), I$_\lambda$(C)~/~I$_\lambda$(A) and I$_\lambda$(DE)~/~I$_\lambda$(A). The $\lambda\lambda$5780 and 5797 DIBs are stronger, by 5\% and 9\%, respectively, towards B than A, as indicated by the residuals in the I$_\lambda$(B)~/~I$_\lambda$(A) plot that precisely match the positions and profiles of the respective DIBs. These two bands are also stronger towards DE than A, but in this case $\lambda$5780 exhibits a much greater change in strength of $\sim$ 20\%, the change for $\lambda$5797 being only \emph{c.} 7\%.  This is a notable finding given that the values of $E_{B-V}$ towards $\rho$~Oph~A and $\rho$~Oph~DE are almost identical with values of 0.47 and 0.48 \citep{pan04}. Despite contamination of the $\rho$~Oph~C spectrum, it is apparent that both $\lambda$5780 and $\lambda$5797 are stronger towards C than A.

\begin{figure}
\centering
\includegraphics[width=0.542\columnwidth]{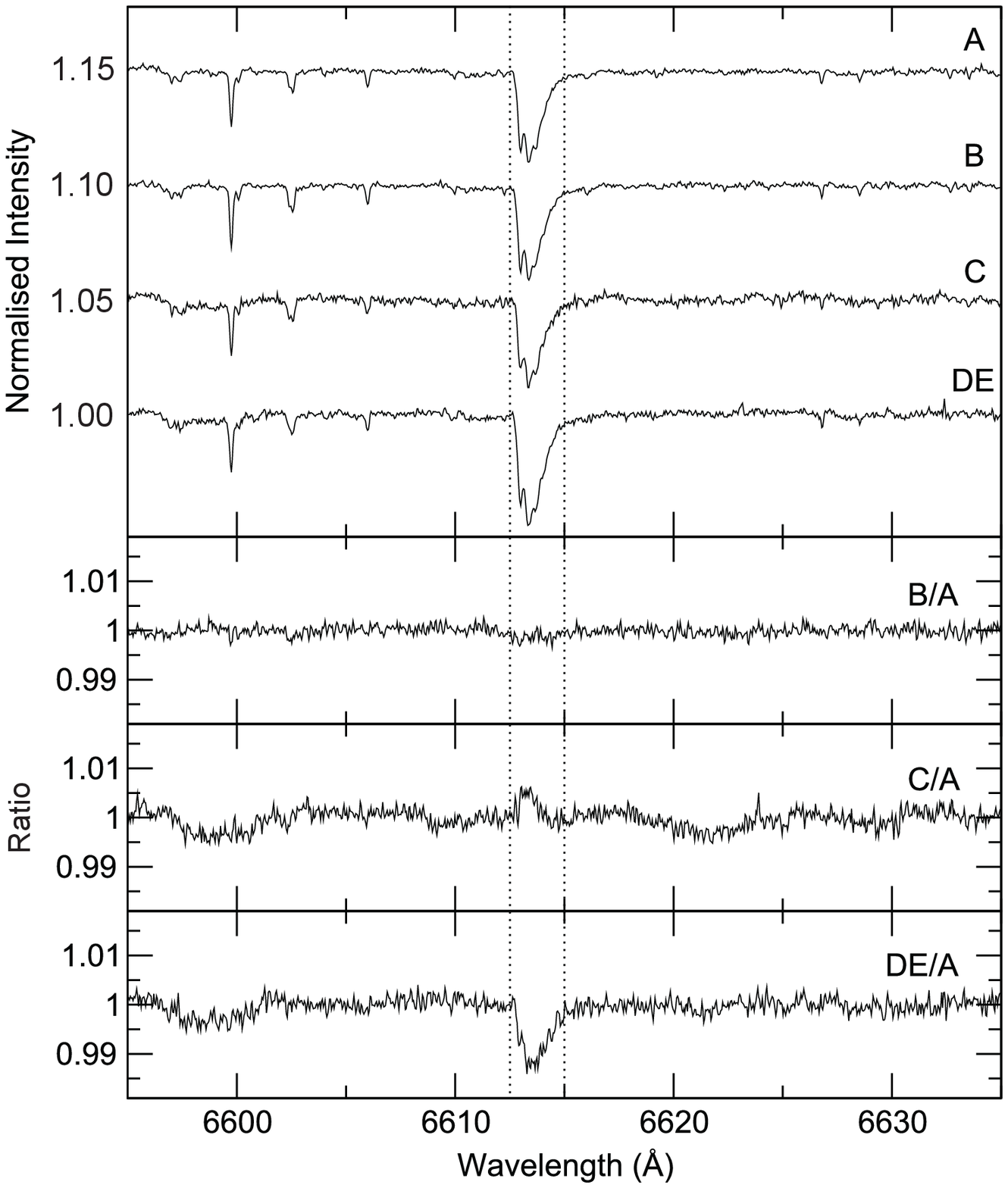}\\
\vspace*{3mm}
\includegraphics[width=0.55\columnwidth]{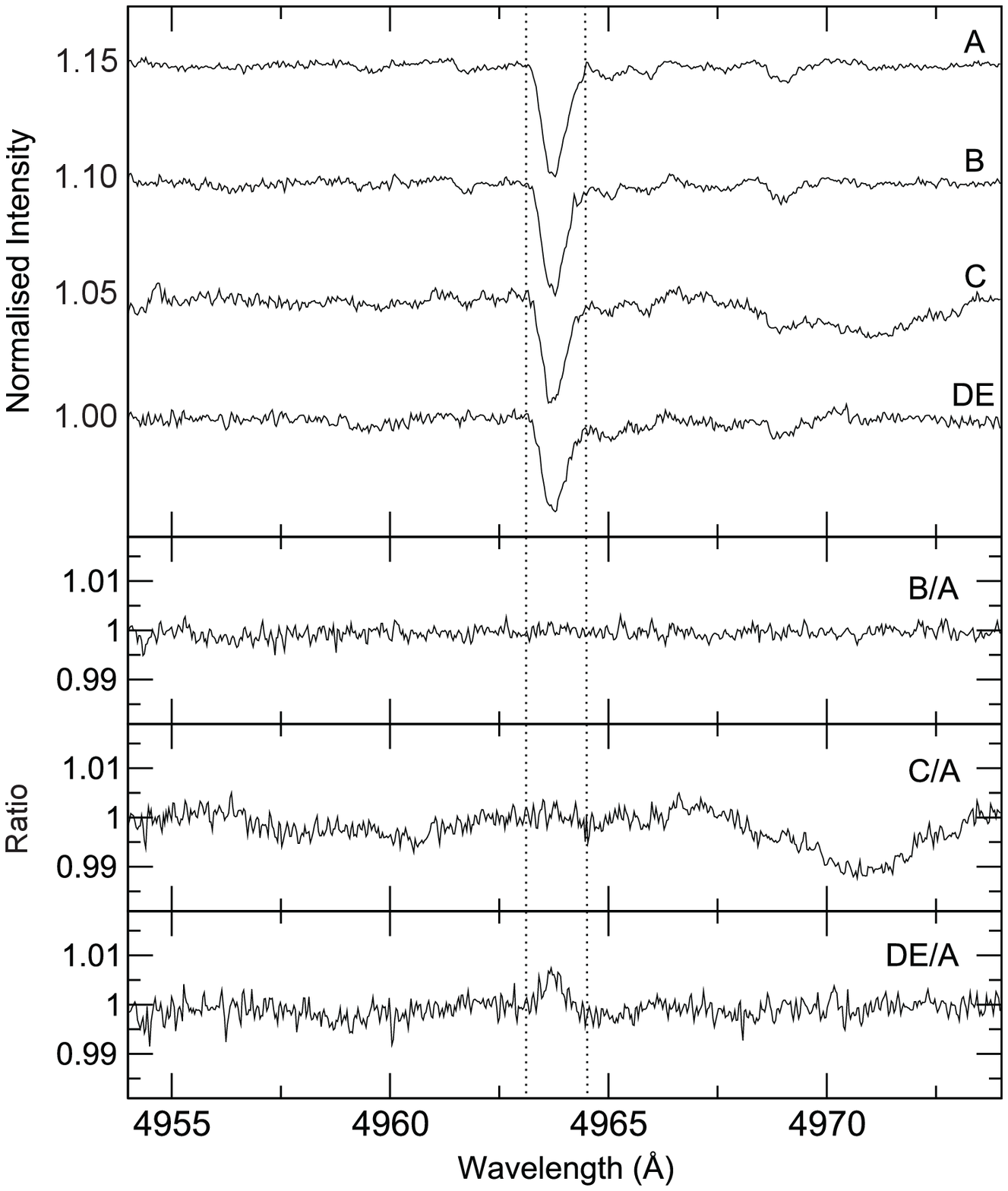}
\caption{Spectra of the region covering the $\lambda$6614 diffuse band (top) and $\lambda$4964 `C${_2}$' diffuse band (bottom). Data are as described in Figure \ref{fig:5780} but with ordinate offsets of 0.05. \label{fig:66144964}}
\end{figure}

Figure \ref{fig:66144964} (top panel) contains the corresponding data for $\lambda$6614 which again show this DIB to be slightly stronger towards $\rho$~Oph~B than A, and much stronger towards DE than A, in a similar fashion to $\lambda\lambda$5797 and 5780. A significant \emph{decrease} in $\lambda$6614 band strength is seen towards $\rho$~Oph C relative to $\rho$~Oph~A. The $\lambda$6614 spectra appear to show some evidence of profile variation which may result from different carrier internal temperatures \citep{cam04} or isotopic enrichments \citep{web96}.

Measurements of the equivalent widths for $\lambda\lambda$5780, 5797, 6614 and seven other diffuse bands are given in Table \ref{tab:W}. Broader DIBs are omitted due to the difficulty in accurately defining their continua in {\'e}chelle spectra and the increased likelihood of overlapping stellar features. The results presented here for the yellow-to-red DIBs (obtained in 2005) match closely the results obtained by \citet{cor06}, who used only the 2004 data (which had slightly lower signal-to-noise, but was observed with better seeing), and focussed on DIBs at wavelengths $>5500$ \AA. The fact that the spectra obtained on these two epochs match so closely despite the different instrumental setups and observing conditions adds confidence to the results presented here.

The $\lambda\lambda$5780, 5797, 5850, 6379 and 6614 DIBs have the greatest central depths of those in our sample, and are the only ones to show a significant difference in strength between $\rho$~Oph~A and B. Small differences were measured for other bands towards these stars, but they are not significant given the noise level. Our data are consistent with all DIBs being slightly stronger towards $\rho$~Oph~B than A (by \emph{c.} 5\%). However, for the AB pair \textit{vs.} C and DE, the situation is more complex. Out of all the $\rho$~Oph sightlines, we find that $\lambda$5850 is strongest towards C, whereas $\lambda$6614 is weakest. Additionally, DE has clearly the strongest $\lambda$6196 DIB, but the weakest $\lambda$4726 (the latter DIB is 40\% weaker towards DE than C and 32\% weaker than AB). This wide variation in behavior of different DIBs between the closely-spaced sightlines in our sample shows that the carriers are being influenced by quite subtle variations in physical/chemical interstellar conditions.

An important advance in diffuse interstellar band studies was the identification of a set of DIBs that have a propensity to follow in strength the column density of diatomic carbon molecules -- the so-called `C${_2}$ DIBs' \citep{tho03}. A spectroscopic property of the C${_2}$ DIBs (possibly related to the structure of their carriers), is that they fall predominantly at the blue end of the spectrum ($\lesssim5500$ \AA), whereas the non-C$_2$ DIBs that we have measured fall at the red end ($>5500$ \AA).  Among the strongest of the C${_2}$ DIBs is $\lambda$4964, data for which are shown in Figure \ref{fig:66144964} (bottom panel). This DIB shows no detectable difference between $\rho$~Oph~A and B but is found to be weaker towards $\rho$~Oph~DE than $\rho$~Oph~A. Interestingly, we find that a total of nine out of the ten observed C${_2}$ DIBs ($\lambda\lambda$4726, 4734, 4964, 4984, 5176, 5418, 5512, 5541, 5546) are measured to be weaker towards DE than AB \citep{smi06}. The remaining $\lambda$5170 C${_2}$ DIB is overlapped by a broad spectral feature, probably stellar in origin, which precludes a proper measurement of its variation. Ten out of eleven of the non-C$_2$ DIBs we observed in the yellow-to-red part of the spectrum (for which measured variations are larger than the errors), show the opposite trend, and are stronger towards DE than AB \citep[data for $\lambda\lambda$5705, 6660 and 7224 are given by][]{smi06}, the single exception being $\lambda5850$. These results strongly support the association of the C$_2$ DIBs as a special group.

\subsection{The ISM towards the $\rho$ Oph star system}

\citet{sno08} studied the atomic (K\,{\sc i}, Na\,{\sc i} and Ca\,{\sc i}) column densities towards $\rho$~Oph~A, C and D and determined the three-dimensional locations of these stars with respect to the $\rho$~Oph cloud complex. They conclude that while $\rho$ Oph AB and D are probably in front of the dense molecular cloud and probe a more diffuse portion, $\rho$~Oph~C more likely lies embedded in a CN-rich molecular part. As discussed by Snow et al. (2008), the ratio of total-to-selective extinction ($R_V$) is unusually high towards $\rho$~Oph, and the far-UV rise is quite flat. These observations are consistent with the presence of relatively large dust grains and an absence of the very smallest grains \citep{sea95}. Notably, the reported $E_{B-V}$ values for the $\rho$~Oph stars are almost identical (Table \ref{tab:stars}), which, given the magnitude of the observed DIB variations, is consistent with previous findings that the DIBs do not arise directly from the large grains that cause optical extinction \citep{sar06}.

\subsection{Comparison of DIBs with known atoms and molecules}

A body of data for atoms and small molecules towards the $\rho$~Oph stars is given in Table \ref{tab:W}, where the column densities quoted are in general the sum over a number of blended velocity components. Unfortunately, the widths of the DIBs preclude their association with any particular component. We measured equivalent widths for 23 lines of C$_2$ in each of our spectra (in the range $7700-8900$ \AA, which covers the (2-0) and (3-0) C$_2$ vibrational bands). Total C$_2$ column densities were calculated using the oscillator strengths of \citet{bak97}; the best-fitting rotational temperatures ($T_ {\rm rot}$(C$_2$)) were obtained using B.~McCall's C$_2$ calculator (at http://dib.uiuc.edu/c2/), and are given in Table \ref{tab:W}. All four lines of sight show absorption due to CN at around $2$~km\,s$^{-1}$, with a common Doppler $b\approx0.8$~km\,s$^{-1}$.

Apart from CH$^+$ which has a larger total column density towards $\rho$~Oph A than B, the errors on the total column densities are too large to permit any clear distinction to be drawn between the properties of the gas towards $\rho$~Oph A and B. However, if one considers only the main $2$~km\,s$^{-1}$ component, CH, CH$^+$, K\,{\sc i} and Ca\,{\sc ii} are all 10-20\% stronger towards B than A.  Our high signal-to-noise K\,{\sc i} equivalent width measurements also show more neutral potassium towards B than A. We infer that the column of DIB carriers increases towards $\rho$~Oph~B in concert with the species noted above. It is difficult, however, to know whether this is due to variations in the gas density, lower atomic (and DIB) depletions, changes in ionization equilibrium, or due to the geometric structure of the cloud.

Our observations towards $\rho$~Oph~C, as well as those of \citet{pan04} and \citet{sno08} show a relative under-abundance of K\,{\sc i} and Ca\,{\sc i} compared with the other three sightlines. Consistent with its ($\sim3$ times) greater CN column density, \citet{sno08} concluded that $\rho$~Oph~C is more deeply embedded in the dense, CN-rich material of the background (star-forming) molecular cloud. This relatively high density material should result in more depletion, which is the most likely explanation for the low K\,{\sc i} and Ca\,{\sc i} column densities.  The lower $\lambda$6614 strength towards $\rho$~Oph~C may also be related to this depletion. The fact that the variations in the `red' DIB strengths are only on the order of a few percent for C compared with AB and DE shows that their carriers are, in general, not closely associated with CN. This result is in accordance with the study of $\lambda\lambda$5797 and 5780 by \citet{wes08}.  By contrast, the `blue' $\lambda$4726 DIB is very strong towards C, which confirms the association of its carrier with molecular material traced by C$_2$ \citep{tho03}.

The greater strengths measured for $\lambda\lambda$5797 and 5850 towards $\rho$~Oph~C are consistent with the assignment of these DIBs to \citet{kre87}'s group III, which is a family of diffuse bands that tend to have relatively narrow profiles. These `KWIII' DIBs (in contrast to those in the KWII group) are generally observed to be strongest in better shielded, denser diffuse clouds where the UV radiation field is weaker and small molecules are more abundant \citep{cam97}. 

The sightline towards $\rho$~Oph~DE is relatively more diffuse than C, as indicated by the lower CN column density. It has a lower depletion factor ($F\ast$; \citealt{jen09}) than A, indicative of a lower gas density. The 2~km\,s$^{-1}$ component towards DE also has the lowest modeled density \citep{pan05}, consistent with it being the more diffuse sightline. The greater C$_2$ rotational temperature measured towards DE may reflect the increased (photo-electric) heating expected to result from a stronger UV radiation field pervading this more diffuse gas. As shown in Figure \ref{fig:5780}, the amount by which the $\lambda$5780 DIB strength increases between A and DE ($\sim20$\%), is greater than the amount by which $\lambda$5797 increases ($\sim10$\%). This pattern is consistent with the hypothesis \citep{kre87,cam97} that the carrier of the $\lambda$5780 DIB favors more diffuse environments than $\lambda$5797. The strengths of other DIBs in the same (KWII) group as $\lambda5780$ ($\lambda\lambda$6196, 6203 and 6614) show a similarly large enhancement ($\sim20$\%) towards DE compared to A/B. With the striking exception of CH$^+$ ($-26$\%), CH, Ca\,{\sc i}, K\,{\sc i} and Ca\,{\sc ii} show increases in column density in the range 15-30\% (for the main 2 km\,s$^{-1}$ component) from A to DE, which is analogous to their increase from A to B discussed earlier, and could plausibly be related to a lower degree of depletion in the gas towards DE.

The relative weakness of the C$_2$ DIBs towards DE is consistent with this being the more diffuse (and therefore, more molecule-poor) sightline of our sample. Our C$_2$ spectra are, however, of insufficient signal-to-noise to discern differences in the C$_2$ column densities between the $\rho$~Oph sightlines; dedicated future C$_2$ observations will be required to confirm this.

\section{Conclusions}

Using high signal-to-noise {\'e}chelle spectra of the stars $\rho$~Oph A, B, C and DE, the strengths of some of the observed diffuse interstellar bands are found to differ by about 5\% on sky-projected distance scales less than \emph{c.} 344 AU. The observed DIB strength variations between $\rho$~Oph A and B are attributed to an increase in the abundances of their carriers/chemical precursors from A to B. Towards $\rho$~Oph C and DE, the DIB behavior is more complex, probably as a result of the greater differences in physical and chemical conditions in the ISM over the relatively larger distances between these sightlines and the different locations of these stars with respect to the background molecular cloud.  The yellow/red DIBs are generally found to be strongest in the relatively diffuse, less-depleted gas towards $\rho$~Oph DE. This is in striking contrast to the C$_2$ (blue) DIBs, which are consistently weaker towards DE than the other stars. The $\lambda$5850 DIB is found to be strongest in the more dense, depleted and molecule-rich diffuse gas towards $\rho$~Oph C, consistent with its assignment to \citet{kre87}'s group III.

Observed DIB strength variations are, in some cases, significantly greater in magnitude than the corresponding variations among atomic and (most) diatomic species; for example the equivalent widths of $\lambda\lambda$4726, 5780 and 6614 towards DE compared with A/B exceed the magnitude of variation in all observed species apart from CH$^+$.  Such variations imply that diffuse band carrier abundances are very sensitive to variations in cloud conditions. Given the small distances between our observed sightlines, differences between the elemental compositions of their gases are expected to be negligible. Variations in diffuse band strengths and atomic and molecular column densities could therefore arise as a consequence of density inhomogeneity or the strength of the incident radiation field.  Recent observations of CH$^+$ and SH$^+$ in the diffuse ISM have highlighted the importance of turbulent dissipation arising from shocks or velocity shears \citep{god09,god12}; this may also play a role in determining diffuse band carrier abundances on small distance scales, particularly if the carrier formation is driven by warm chemistry or through release of material from interstellar dust grains.

A more complete explanation for the origin of the observed small-scale variations in DIB strengths in the $\rho$~Oph system will require a detailed knowledge of the properties of each sightline, including the density, temperature, (gas and solid-phase) chemical abundances and radiation field. This could be achieved through high signal-to-noise, high-resolution optical/UV/IR studies of atomic and molecular absorption lines including H\,{\sc i}, H$_2$, C, C$_2$, and other neutral and ionic species sensitive to physical conditions in the ISM.

\acknowledgements
We thank PATT for the allocation of AAT time and T \& S, EPSRC for studentships, and STFC for a visitor grant. MAC thanks the NASA Astrobiology Institute through The Goddard Center for Astrobiology. SJF thanks M. M. Dworetsky and I. D. Howarth for discussions on 
photospheric line-profile variations in $\rho$~Oph C.

\begin{deluxetable}{cclccc}
\tablecaption{Program star properties\label{tab:stars}}
\tablewidth{0pt}
\tablehead{
\colhead{Star}&\colhead{HD}&\colhead{Sp. Type}&\colhead{$V$}&\colhead{$E_{B-V}$}&\colhead{S/N (6000 \AA)}}
\startdata
$\rho$~Oph A&147933&B2/B3 V&5.0&0.47&1200\\
$\rho$~Oph B&147934&B2 V&5.9&0.47&1100\\
$\rho$~Oph C&147932&B5 V&7.3&0.47&800\\
$\rho$~Oph D&147888&B3/B4 V&6.8&0.48&900\\
\enddata
\tablecomments{Spectral types are from the Michigan Catalogue of Spectral Types \citep{hou88}. Reddening values ($E_{B-V}$) are from \citet{pan04}.}
\end{deluxetable}

\begin{deluxetable}{lrrrrr}
\tablecaption{$\rho$~Oph diffuse interstellar band and atomic/molecular line measurements \label{tab:W}}
\tablewidth{0pt}
\tablehead{&\colhead{A}&\colhead{B}&\colhead{C}&\colhead{DE}&\colhead{Error}}
\startdata
&\multicolumn{4}{c}{Equivalent widths (m\AA)}\\
\hline
$\lambda$4726&99.3&97.8&112.9&67.4&4.3 \\
$\lambda$4964&23.5&22.0&21.5&19.5&0.8\\
$\lambda$5780&188&197&[199]&227&2.8 \\
$\lambda$5797&49.7&54.2&[56.0]&55.2&1.8\\
$\lambda$5850&29.8&31.9&35.2&25.2&1.8\\
$\lambda$6196&15.9&15.6&15.0&18.8&2.0\\
$\lambda$6203&42.0&42.5&43.0&47.2&6.0\\
$\lambda$6376&14.8&15.8&15.5&21.0&1.3\\
$\lambda$6379&27.1&28.3&28.5&34.7&1.3\\
$\lambda$6614&63.9&67.7&62.1&84.9&1.3\\[1.5mm]
K\,{\sc i} $\lambda$7699&88.1&89.5&77.7&90.5&0.3\\
\\
\hline
&\multicolumn{4}{c}{\hspace*{-1cm}Column densities and derived values \hspace*{-2cm}}\\
\hline
$N({\rm H\,\textsc{i}}$)~/~10$^{21}$&4.3&\nodata&\nodata&4.9&1\\
$N({{\rm H}_2}$)~/~10$^{20}$&3.7&\nodata&\nodata&3.0&1\\
$N({\rm K\,\textsc{i}}$)~/~10$^{12}$&\phantom{W}1.03&\phantom{W}1.09&\phantom{W}0.80&\phantom{W}1.10&0.05\\
$N({\rm Ca\,\textsc{i}}$)~/~10$^{10}$&1.74&1.69&1.62&2.02&0.09\\
$N({\rm Ca\,\textsc{ii}}$)~/~10$^{12}$&1.75&1.96&1.92&1.94&0.1\\
$N({\rm CH})$~/~10$^{13}$&2.37&2.33&1.92&2.18&0.1\\
$N({\rm CH}^+)$~/~10$^{13}$&1.59&1.43&0.70&0.75&0.06\\
$N({\rm CN})$~/~10$^{13}$&0.21&0.20&0.60&0.21&0.01\\
$N({\rm C}_2)$~/~10$^{13}$&4.1&2.9&3.3&4.0&1\\
$T_{\rm rot}$(C$_2$) / K&51&49&46&56&1\\
$n$ / cm$^{-3}$&625&450&1100&425&\nodata\\
Depletion ($F\ast$)&1.09&\nodata&\nodata&0.88&0.07\\
\enddata
\tablecomments{Upper part of table gives equivalent widths of interstellar features measured towards $\rho$~Oph stars; values in brackets indicate probable stellar photospheric line contamination.  Lower part of table gives atomic and molecular column densities (in cm$^{-2}$) taken from \citet{pan04} except: C$_2$ (from this work), H~\textsc{i}, H${_2}$ and depletion factor ($F\ast$) from \citet{jen09}. Gas number densities $n$ are from the chemical model of \citet{pan05} and pertain to the CN-containing 2~km\,s$^{-1}$ component only. Final column gives an average ($\pm$) error estimate on the respective values.}
\end{deluxetable}

\end{document}